\documentclass[12pt]{iopart}

\usepackage{graphicx}
\usepackage{epsfig}
\usepackage{iopams}
\begin{document}

\title{Quantum frustration of dissipation by a spin bath}

\author{D D Bhaktavatsala Rao,$^1$ Heiner Kohler,$^2$ and Fernando Sols$^3$}

\address{$^1$ Department of Physics, Indian Institute of Technology Kanpur, Kanpur 208016, India. \\
$^2$ Department of Physics, University of Duisburg-Essen, D-47057 Duisburg, Germany \\
$^3$ Departamento de F\'{\i}sica de Materiales, Universidad
Complutense de Madrid, E-28040 Madrid, Spain}
 \ead{ddbrao@iitk.ac.in}
\begin{abstract}
We investigate the evolution of a central spin coupled to a spin
bath without internal dynamics. We compare the cases where the bath couples to one or two
components of the spin. It is found that the central spin dynamics is
enhanced in the latter case, which may be interpreted as a frustration
of dissipation. However, the quantum purity of the spin decays fast
in both scenarios. We conclude that symmetric coupling of the bath
to two orthogonal components of the spin inhibits dissipation but
not decoherence.

\end{abstract}
\maketitle
\section{Introduction}
 The study of the role of a dissipative environment is of central importance to the field of quantum computation and for the fundamental understanding of the transition from quantum to classical behavior.  In that context, the dissipative two-level system (TLS) is a well-studied paradigm \cite{leg87, wei99}. The generic TLS, which in particular can be a spin ${\small \frac{1}{2}}$ particle, experiences dissipation due to its coupling to a bath of harmonic oscillators. In the resulting spin-bath problem, an external magnetic field interacts with one component of the spin operator (e.g. $S^z$) while a second component (e.g. $S^x$) couples to an oscillator bath. Depending on the relative strength of the two interactions, one can switch over between underdamped and overdamped behaviour, as may become manifest through quantities such as the average spin energy or various correlation functions.  Recently Castro Neto {\it et al.} [3, 4] have shown that if two components ($S^x$ and $S^y$) of the effective spin are coupled to two different baths, then the competing effects of those baths can reduce the effect of dissipation. In particular, they have found that, for a given coupling strength, symmetric coupling to the two spin components is less decoherent than coupling to a single component. Moreover, in the case of symmetric coupling to $S^x$ and $S^y$, coherent behaviour is preserved for arbitrarily strong coupling. These two properties are remarkable because one would naively expect that coupling to a higher number of bath oscillators would increase the effect of dissipation. Rather, they have shown that the competition between the baths contributes to protect the TLS energy gap. The reduction of dissipation stems from the non-commutative character of the spin operators coupled to the different baths. Logically, if the two baths interact with the same spin component, then no dissipation reduction is observed. In Refs. \cite{cas03, nov05}, it is argued that this feature arises from the lack of a preferred basis to which the TLS may relax at long times. This new phenomenon has been coined quantum frustration of decoherence, since it is interpreted as the frustrated attempt of the two environments to ``measure'' simultaneously two non-commuting observables. The study of Refs. \cite{cas03, nov05} has been restricted to equilibrium properties such as the transverse susceptibility, which was evaluated using the numerical renormalization group.

Features of quantum frustration were also reported in Ref. \cite{koh04}, where it was noted that the phase-number variable of a superconducting
Josephson junction is coupled simultaneously to two different dissipative environments: through the phase to the quasiparticle field and through the Cooper pair number to the quantum electromagnetic field. The result is that the uncertainty in the macroscopic phase has contributions from both baths which tend to cancel each other. They never cancel completely because, in that particular case, the sources of dissipation differ widely both in nature and strength. Frustration of decoherence in Josephson networks has also been investigated \cite{giu07}

In Refs.~\cite{koh05, koh06b} the dynamics of an oscillator coupled through its position and momentum to two different
oscillator baths was studied. The problem can be shown to be equivalent to that of a large (quasiclassical) spin impurity in a ferromagnetic environment. It was noted that the dissipative oscillator may be driven from overdamped to underdamped behaviour by the symmetrical addition of a second bath. This surprising effect was found only for the case where the momentum operator ($p$) and the position operator ($q$) are coupled to two different baths.
Like in Refs.~\cite{cas03,nov05}, these effects were investigated through equilibrium properties such as the position-position response function. However, some dynamical aspects were also analyzed, noting that the purity of the quantum oscillator decays faster in the presence of two baths than in the presence of a single bath. This last result is important because it reveals that the competition between two baths coupling to non-commuting observables is not a universal panacea to suppress decoherence. Our present study is motivated by the need for a more detailed understanding of the equilibrium and dynamical properties of a quantum system in the presence of competing environments.

In the studies of quantum frustration made earlier \cite{cas03,nov05,koh04,koh05,koh06b}, the bath has been modeled by a set of non-interacting oscillators. However, if the interpretation is correct that the essence of quantum frustration stems from the canonically conjugate character of the two observables which couple to separate baths, then one should expect a similar behaviour to appear when the dissipative environment is formed by a bath of spins acting on a central spin impurity. We note that, because of the vector nature of the spin, the spin bath can by itself be viewed as formed by several baths. Thus a single spin bath may exhibit features of quantum frustration. Spin baths have been studied \cite{pro00,cuc05,rel07,zur07,ros07a,ros07b,rao07,lai08} as an alternative to the conventional oscillator models of quantum dissipation \cite{leg87,wei99}. They are known to give rise to non-Markovian evolution, with the system evolution showing a strong dependence on the polarization of the initial state \cite{rao06}.

In this work we shall evaluate the dynamical properties of a TLS coupled to a bath of spins. Similar to the work of Novais {\it et.al.} \cite{cas03,nov05}, we shall consider the situation where the components of the spin are coupled to two different baths and study the effects of frustration arising due to the non-commuting nature of the spin components. The work of Refs. \cite{cas03,nov05} was based on the method of the numerical renormalization group. By assuming that the bath has no internal dynamics of its own, we are able to perform an analytical study.  Comparing the case of coupling to a single component to that of symmetric coupling to both components, we find that the spectral function behaves similarly to the study of Refs. \cite{cas03,nov05}. Namely, the spectral function develops a peak in the symmetric case which is absent in the case of single-component coupling. In Refs. \cite{cas03,nov05}, this was interpreted as the preservation of decoherence arising from the frustrated attempt of the two environments to measure two non-commuting observables. However, we find that the emergence of the peak as the second bath intervenes is compatible with a fast decay of the quantum purity of the central spin. This means that the frustration induced by the two competing environments has more to do with dissipation than with decoherence. Energy relaxation is indeed inhibited by the presence of a second bath coupled to the other spin component, while the quantum purity decays fast in the presence of a second bath, with only a minor form of frustration revealed by a short-lived revival which will be discussed.

Section 2 is devoted to the presentation of the model of a central spin coupled to a spin bath. In Sec. 3 we study the time evolution analytically, deriving expressions for the expectation values and the quantum purity of the central spin. Sections 4  and 5 focus on the density of states and the response function of the central spin. Section 6 deals with the tailoring of the spin bath properties which mimics the behavior of a spin coupled to a conventional bath of harmonic oscillators. Finally, the main conclusions of this work are summarized and discussed in Sec. 7.


\section{Central spin model}

We consider the dynamics of a two-level-system which is linearly coupled through two non-commuting observables to two independent environments of two-level-systems. We will refer indistinctively to both the central impurity and the constituents of the bath as particles of spin ${\small \frac{1}{2}}$ or two-level-systems. If the spins of the environment carry their own dynamics, in general there are no conserved quantities other than energy and the system can not be treated analytically without approximations.  However, in most of the solid-state spin systems where the spin bath interaction is a dominant source of mechanism for the dissipation of a TLS, the internal bath dynamics is generally very slow (for example in quantum dot systems, where the bath spins are nuclear spin half particles and the TLS is the electronic spin, \cite{loss}). We therefore assume that both environments carry no dynamics of its own, i.~e.~that their Hamiltonians are zero. Thus the bath dynamics is
  exclusively due to its interaction with the central spin \cite{rao07}.
The total Hamiltonian of system and bath is then given by
\begin{eqnarray}
\label{fham0}
H = H_{\rm S} + H_{\rm SB}
\end{eqnarray}
where
\begin{eqnarray}
\label{fham}
H_{\rm S} &=& \omega_0 S^z,  \\
H_{\rm SB} &=& g_1 S^x \sum_{k=1}^N I^x_k + g_2 S^y \sum_{l=1}^N J^y_l,
\end{eqnarray}
where $S^i$, $i=x,y,z$ are the components of the spin operator of the central spin and $I^x_k$ and $J^y_k$ are spin operators of the bath spins. We assume homogeneous interaction between the central spin and the baths. Moreover we assume that the number of spins of each environment to be the same. The strength of the coupling to each environment is thus described by one parameter $g_i$ ($i=1,2$) only. For a single bath--environment the case of non--homogeneous coupling ($g_1,g_2$ dependent on index $k,l$, respectively) was solved explicitly by in Ref. \cite{cuc05}. However, it was shown in Refs. \cite{cuc05, rao07} that the only effect of inhomogeneous interaction is that of destroying certain revival effects. All other features can be captured within the homogeneous interaction between system and bath. The main advantage of the homogeneous interaction approximation is that exact, closed-form expressions for the expectation values and correlation functions can be obtained.

The Hamiltonian (\ref{fham0}) is similar to that employed in Refs. \cite{cas03,nov05} in that two different environments couple to the two perpendicular components of the central spin. The main difference with the model of Refs. \cite{cas03,nov05} is the non-dynamic character of the bath, which we consider, which contrasts with the oscillator bath there considered. We shall examine whether our simpler model (\ref{fham0}) can yield frustration effects similar to those obtained from the more complex model of Refs. \cite{cas03,nov05}, which was solved with the numerical renormalization group method.

In the particular case where the spin bath couples to only one component of the central spin, a number of non-trivial effects are known to appear, despite its apparent simplicity. One instance is the crossover from overdamped to underdamped behaviour as the coupling strength increases, similar to the spin-boson model where the TLS is coupled to an oscillator bath \cite{leg87,wei99}. For a more complete account on the dynamics of a spin coupled to a single bath we refer to Refs. \cite{pro00,rao07}.

\section{Time evolution}
\label{secgen}
In the absence of bath dynamics the $x$-- and $y$--components of the total spin of the respective environments, $I_{\rm tot}^x=\sum_{k} I_k^x$ and $J_{\rm tot}^y=\sum_{k} J_k^y$, are conserved quantities. We write the total Hilbert space as a tensor product ${\cal H}_{\rm S}\otimes{\cal H}_{\rm B}$. The total $2^{2N}$ dimensional Hilbert space of the baths ${\cal H}_{\rm B}$ decomposes into invariant subspaces ${\cal H}^{(m_1,m_2)}_{\rm B}$. These are labeled by the eigenvalues $m_1$, $m_2$ of the
total spin operators $I_{\rm tot}^x$ and $J_{\rm tot}^y$. Each $m_i$ ($i=1,2$) runs from  $-N/2$ to $N/2$. The Hamiltonian (\ref{fham0}) acts on the subspace  ${\cal H}_{\rm S}\otimes{\cal H}^{(m_1,m_2)}_{\rm B}$ as
\begin{eqnarray}
H |\sigma\rangle|\alpha_1,\alpha_2;m_1,m_2\rangle & = & (\omega_0 S^z + S^x m_1 g_1 +  S^y m_2 g_2) |\sigma\rangle|\alpha_1,\alpha_2;m_1,m_2\rangle\ ,
\end{eqnarray}
for $|\alpha_1,\alpha_2;m_1,m_2\rangle$ $ \in {\cal H}^{(m_1,m_2)}_{\rm B}$ and $|\sigma\rangle \in {\cal H}_{\rm S}$. Here $\alpha_i$, which is not important in the following, labels the irreducible representation. We therefore can write $H$ most conveniently as a direct sum $H = \oplus_{m_1,m_2=-N/2}^{N/2} H_{m_1m_2}$ where
\begin{eqnarray}
\label{heff}
H_{m_1m_2} & = & (\omega_0 S^z + g_1 m_1 S^x + g_2 m_2 S^y)
                 \otimes {\mathbb I}_{\lambda_{m_1}}\otimes {\mathbb I}_{\lambda_{m_2}}\nonumber\\
           & = &  \vec{S}\cdot\vec{\Omega}_{m_1m_2} \otimes {\mathbb I}_{\lambda_{m_1}}\otimes {\mathbb I}_{\lambda_{m_2}} \ .
\end{eqnarray}
Here $\mathbb{I}_{\lambda}$ is the $\lambda\times \lambda$ unit matrix and the parameters $\lambda_{m_i}$,
\begin{eqnarray}
\label{lambda}
\lambda_{m_i} & = &
\left(
  \begin{array}{c}
    N \\
    N/2-m_{i} \\
  \end{array}
\right)
\ ,\qquad i=1,2
 \end{eqnarray}
measure the dimension of the invariant subspace ${\cal H}^{(m_1,m_2)}_{\rm B}$. From Eq.~(\ref{heff}) it is clear that the effect of the environment is to give rise to an effective magnetic field $\vec{\Omega}_{m_1m_2}=(m_1g_1,m_2g_2,\omega_0)$. However, this effective magnetic field is different from the static magnetic field pointing in the $z$--direction, since it does not take a single value but rather is a distribution characterized by the degeneracy coefficients $\lambda_{m_i}$.

Since $H_{m_1m_2}$ acts on the subspace of the environment, we will often write $ H_{m_1m_2}= \vec{S}\cdot \vec{\Omega}_{m_1m_2}$ for short. The eigenvalues of $H_{m_1m_2}$ are $\pm \Omega_{m_1m_2}/2$, where we introduced the frequency $\Omega_{m_1m_2}
= (\omega_0^2 + g_1^2m_1^2 + g_2^2m_2^2)^{1/2}$. We denote the eigenstates of $H_{m_1m_2}$ by $|\pm,m_1,m_2\rangle$. They are related to the eigenstates $|\uparrow \rangle$, $|\downarrow\rangle$ of the non--interacting system Hamiltonian $H_{\rm S}$ by the unitary transformation
\begin{eqnarray}
\label{trafo}
\hspace{-2cm}
\left(\begin{array}{c}
|+,m_1,m_2\rangle\cr |-,m_1,m_2\rangle
\end{array}\right)\
&=&  \left(\begin{array}{cc}
   \cos\theta_{m_1m_2}&\sin\theta_{m_1m_2} e^{i\phi_{m_1m_2}}\cr
 - \sin \theta_{m_1m_2}e^{-i\phi_{m_1m_2}}&\cos\theta_{m_1m_2}
\end{array}\right)
\left(\begin{array}{c}
|\uparrow \rangle\cr |\downarrow \rangle
\end{array}\right)
\end{eqnarray}
where the angles $\phi_{m_1m_2}$ and $\theta_{m_1m_2}$ are given by
\begin{eqnarray}
\phi_{m_1m_2} &=& \arctan \left( \frac{m_2 g_2}{m_1g_1} \right) \\
\cos^2\theta_{m_1m_2} &=& \frac{\Omega_{m_1m_2}+\omega_0}{2\Omega_{m_1m_2}}\ .
\end{eqnarray}
The case of a single environment is recovered by setting $g_2$ or, equivalently, $\phi_{m_1m_2}$ equal to zero.

The time evolution operator is straightforwardly derived from Eq.~(\ref{heff}). We obtain
$U = \oplus_{m_1,m_2=-N/2}^{N/2} U_{m_1m_2}$ with
\begin{eqnarray}
\label{timeevol}
U_{m_1m_2}(t)& = &\cos\left(\frac{t}{2} \Omega_{m_1m_2}\right) ~\mathbb{I}_2 + 2i\frac{\sin\left(\frac{t}{2} \Omega_{m_1m_2}\right)}{\Omega_{m_1m_2}} \vec{S}\cdot\vec{\Omega}_{m_1m_2} \ ,
\end{eqnarray}
We can decompose an arbitrary system (central spin) operator $O$ as $O=\oplus_{m_1,m_2=-N/2}^{N/2} O_{m_1m_2}$. In particular we are interested in the Heisenberg spin operator $\vec{S}(t)$ and its commutators and anticommutators at different times. Using (\ref{timeevol}), we find
\begin{eqnarray}
\vec{S}_{m_1m_2}(t) & = & \cos(\Omega_{m_1m_2}t)\vec{S}(0) -
\sin(\Omega_{m_1m_2}t)(\vec{S}(0)\times\vec{n}_{m_1m_2}) \nonumber\\
                    &&\quad + [1-\cos(\Omega_{m_1m_2}t)](\vec{S}(0)\cdot\vec{n}_{m_1m_2})\vec{n}_{m_1m_2}\label{exp}\\
-i [S_{m_1m_2}^i(t),S_{m_1m_2}^j(0)] & =& \cos(\Omega_{m_1m_2}t) \epsilon_{ijk} S^k(0) \label{corr1}\\
                &&\ + \sin(\Omega_{m_1m_2}t)\left(\delta_{ij}\vec{n}_{m_1m_2}\vec{S}(0)-S^i(0) n^j_{m_1m_2}\right)\nonumber\\
                  &&\quad +[1-\cos(\Omega_{m_1m_2}t)]n^i_{m_1m_2}\left(\vec{S}(0)\times \vec{n}_{m_1m_2}\right)_j \nonumber\\
2 \{S_{m_1m_2}^i(t),S_{m_1m_2}^j(0)\}& = &  \cos(\Omega_{m_1m_2}t) \delta_{ij}
                                            -\sin(\Omega_{m_1m_2}t) \epsilon_{ijk}n^k_{m_1m_2}\nonumber\\
                                        &&\quad + \left[1- \cos(\Omega_{m_1m_2}t)\right] n^i_{m_1m_2}n^j_{m_1m_2} \ .\label{corr2}
\end{eqnarray}
The vector $\vec{n}_{m_1m_2}$ is a unit vector pointing in the direction of the effective magnetic field $\vec{\Omega}_{m_1m_2}$.

If the density matrix $\rho(t)$ of the total system is initially invariant under rotations within a subspace ${\cal H}^{(m_1,m_2)}_{\rm B}$ due to the trivial action of the Hamiltonian in this subspace, this invariance will persist at all times. In particular, the density matrix can be written for all times as  $\rho(t)=\oplus_{m_1,m_2=-N/2}^{N/2} \rho_{m_1m_2}(t)$. This means that $\rho(t)$ shares for all times the block structure of the Hamiltonian.
If $\rho(0)$ fulfills this condition, the expectation value of an arbitrary system operator $O(t)$ with respect to $\rho(0)$ can then be written as
\begin{eqnarray}
\langle O(t) \rangle & \equiv & {\rm tr} \left[\rho(0) O(t)\right]\nonumber\\
                             &=& \sum_{m_1,m_2=-N/2}^{N/2}\lambda_{m_1}\lambda_{m_2}
                                O_{m_1m_2}(t)\rho_{m_1m_2}(0) \label{expgen1}
\end{eqnarray}
In the following we will analyse the expectation values of the operators (\ref{exp}) to (\ref{corr2}) with respect to an initially unpolarized bath. Since the magnetic field applied along the $z$--direction only affects the central spin, we take this to be initially in the ground state determined by $H_{\rm S}$ and consequently choose the initial density matrix as
\begin{equation}
\label{rhounpol}
\rho(0) \ =\ \frac{1}{2^{2N}}|\uparrow\,\rangle\langle\, \uparrow|\otimes {\mathbb I}_{2^N}\otimes {\mathbb I}_{2^N}\ .
\end{equation}
We immediately see that, in this state, $\langle S^i(0) \rangle = -\delta_{i3}/2$.

Using the above formalism for evaluating the dynamical properties of the TLS we shall now calculate various quantities for the TLS operators and study the effects brought about by the coupling to two different baths.

\begin{figure}[tbp]
\begin{center}
   \includegraphics[width=10.0cm]{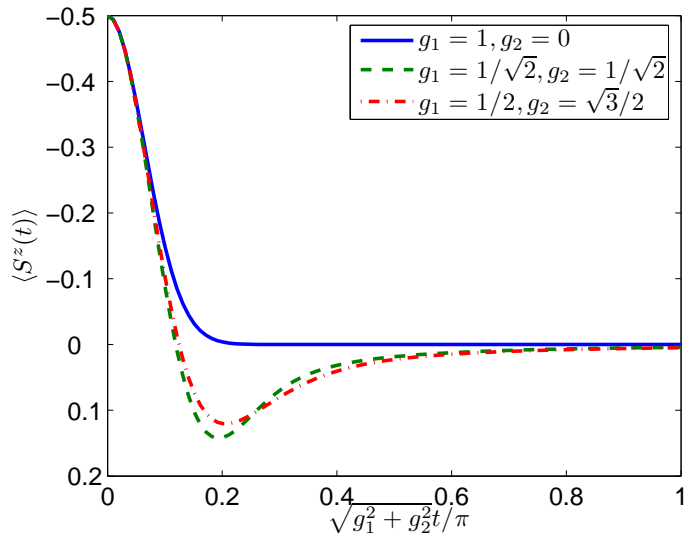}
\end{center}
 \caption{ The magnetization of the TLS is plotted with time for various values of the system bath interaction, in zero-field.
In the presence of two different baths a change of sign arises during the time evolution. The sign reversal reaches its maximum value when the two baths are identical. The total number of spins in each bath is taken to be $N = 100$. The system bath couplings given in the inset are dimensionless ($g_i/\sqrt{g_1^2+g_2^2}$).}\label{expzerofield}
 \end{figure}

\subsection{Expectation values}

The expectation values $\langle S^i(t)\rangle$ can now be calculated using Eqs.~(\ref{exp}) and (\ref{rhounpol}). We find
\begin{eqnarray}
\label{expresult}
\langle S^z(t)\rangle & = & \frac{-1}{2^{2N+1}}\sum_{m_1,m_2=-N/2}^{N/2}
   \lambda_{m_1}\lambda_{m_2}\nonumber\\
&&\quad \left(\frac{g_1^2m_1^2+g_2^2m_2^2}{\Omega_{m_1m_2}^2}\cos(\Omega_{m_1m_2}t)+ \frac{\omega_0^2}{\Omega^2_{m_1m_2}}\right)\\
\langle S^x(t)\rangle  &=&   \langle S^y(t)\rangle \ =\ 0
\end{eqnarray}
In the absence of external field $\omega_0 = 0$, we would expect that the initial polarization of the TLS would decay faster in comparison to the single bath case, since the total number of spins with which the TLS is interacting is doubled.
In Fig.~\ref{expzerofield} we have plotted the time variation of the $\langle S^z(t) \rangle$ for various values of $g_1, g_2$ keeping
$g = \sqrt{g_1^2+g_2^2}$ constant. As one can see, for a single bath the polarization decays to zero very fast, whereas in the presence of the second bath, the decay is comparatively slow. In contrast to the case of single bath, one observes a change of sign in the time-dependent behaviour of the polarization of the TLS, indicating the presence of a nonzero field.

\begin{figure}[tbp]
\begin{center}
   \includegraphics[width=10.0cm]{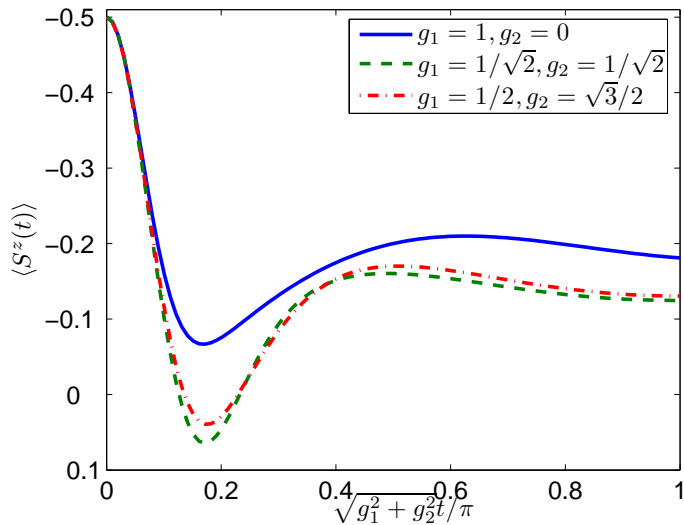}
\end{center}
 \caption{Same as Fig. \ref{expzerofield} but for a nonzero field such that $\omega_0/\sqrt{g_1^2+g_2^2} = 2$. In the presence of the external field the magnetization for the single bath case decays non-monotonically and saturates to a non-zero value.}\label{expfinitefield}
 \end{figure}

In Fig.~\ref{expfinitefield} the time variation of the $\langle S^z(t) \rangle$ is plotted for non--vanishing external field $\omega_0\neq 0$. We find that the polarization saturates to a finite value at long times with faster oscillations in the case of a
symmetric double bath, which is consistent with the spectral properties discussed later in the text.
Inspection of Figs. \ref{expzerofield} and \ref{expfinitefield} reveals that the change of sign occurs in the single bath case only if there is a nonzero field, while it is observed for both zero and nonzero field in the case of a symmetric double bath. Since the baths are completely unpolarized (peaked at $m=0$), it is clear that the effective field responsible for this change of sign can only stem from the competing effect of two baths coupled to non-commuting components of the central spin.

\subsection{Quantum purity}

For an arbitrary density matrix $\rho$, purity is defined as ${\cal P}= {\rm Tr} \rho^2$.
Purity is a convenient, basis-independent measure of the degree of coherence, if $\rho$ is the reduced density matrix of the central spin (that which results from tracing out the bath degrees of freedom in the total density matrix). In our case the decay of purity is directly related to the relaxation of the spin expectation values to equilibrium.
\begin{equation}
{\cal P}(t) \ =\ \frac{1}{2} +\sum_{i=1}^3 \langle S^i(t)\rangle^2 .
\end{equation}

\begin{figure}[tbp]
\begin{center}
   \includegraphics[width=10.0cm]{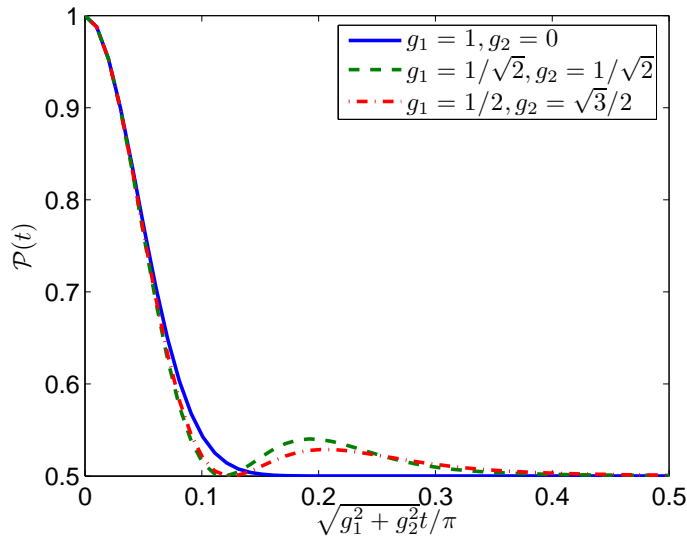}
\end{center}
 \caption{ The purity of TLS is plotted against time for various values of the system bath interaction. Though the initial decay rate strongly depends on the combined interaction strength $\sqrt{g_1^2+g_2^2}$ (which is same for all the three cases), the TLS looses its initial polarization faster in the presence of a single bath in comparison to the case of two baths. The external field is set to zero and the number of spins in each bath are taken to be equal with $N = 100$. The system bath couplings given in the inset are dimensionless ($g_i/\sqrt{g_1^2+g_2^2}$).} \label{purity}
 \end{figure}

The result is plotted in Fig. \ref{purity}. We notice that, both in the single and double symmetric bath cases, the purity decays fast to its minimum value 1/2. In the symmetric case, we notice a small, short-lived revival that may be interpreted as a weak form of decoherence frustration which however does not affect the long time behavior of the central spin.

\section {Density of states}

\begin{figure}[tbp]
\begin{center}
   \includegraphics[width=10.0cm]{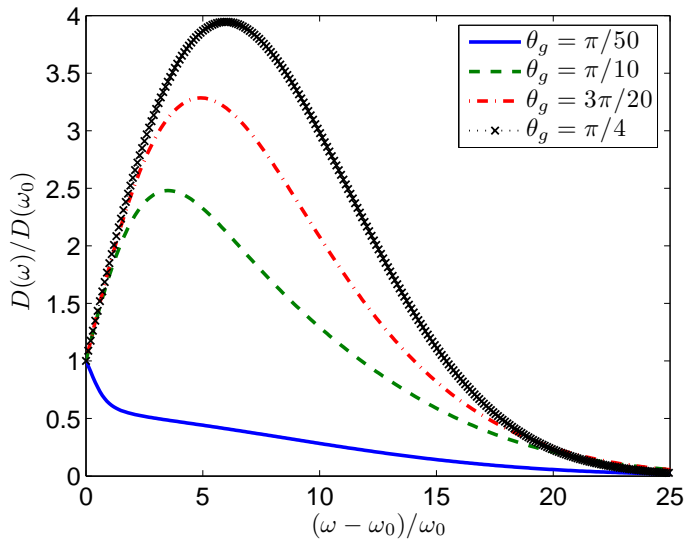}
\end{center}
 \caption{Density of states as a function of frequency in normalized units.}\label{DOS}
 \end{figure}

To get further analytical insight we introduce the function
\begin{eqnarray}
\label{densstate}
D(\omega)&=& \frac{1}{2^{2N+1}}\sum_{m_1,m_2=-N/2}^{-N/2}\lambda_{m_1}\lambda_{m_2}
           \left[\delta(\omega-\Omega_{m_1m_2})+ \delta(\omega+\Omega_{m_1m_2})\right] \
\end{eqnarray}
which is essentially the density of states, normalised to fulfill the sum rule $\int d\omega D(\omega) =1$. The expectation value $\langle S^z(t)\rangle$ is related to $D(\omega)$ by
\begin{equation}
\label{ct}
\langle S^z(t)\rangle \ =\ -\frac{1}{2}\int d\omega D(\omega)\left(\cos(\omega t) \frac{\omega^2-\omega_0^2}{\omega^2} +
                              \frac{\omega_0^2}{\omega^2}\right)  \ .
\end{equation}
$D(\omega)$ can further be evaluated by using the approximation
\begin{equation}
\label{c}
 \frac{1}{2^N}\sum_{m} \lambda_{m} \approx \sqrt{\frac{2}{\pi N}}\int_{-\infty}^{\infty}dm{\rm e}^{-2m^2/N} \ ,
\end{equation}
for the binomials $\lambda_{m_i}$, which is known as Laplace--de Moivre formula in probability theory \cite{gri01} and which is valid only for large $N$. We find
\begin{eqnarray}
 \label{densstate1}
  D(\omega) &=& \frac{4|\omega|\theta(\omega-\omega_0)}{N g^2 \sin (2\theta_g})
I_0\left(\frac{4\cot(2\theta_g)(\omega^2-\omega_0^2)}{Ng^2\sin(2\theta_g)}\right)
 \exp \left(-\frac{4(\omega^2-\omega_0^2)}{Ng^2\sin^2(2\theta_g)}\right)\ .
 \end{eqnarray}
where the total coupling strength $g$ and the angle $\theta_g$ are defined by
\begin{eqnarray}
\label{theta}
g_1 & = & g \cos\theta_g \nonumber\\
g_2  & = & g \sin\theta_g  \ ,
\end{eqnarray}
and $I_0$ is the modified Bessel function. The case $\theta_g = 0$ corresponds to the single bath and the case $\theta_g = \pi/4$ corresponds to two identical baths.
In Eq.~(\ref{densstate1}) the limit of a single bath can be taken by using the asymptotic expansion of the modified Bessel function
$\lim_{z\to\infty} I_0(z)= e^z/\sqrt{2\pi z}+\ldots$ The results are plotted in Fig. \ref{DOS}.
As the coupling of the TLS with the baths becomes symmetric, i.e. $\theta_g \rightarrow \pi/4$, the density of states peaks at a frequency away from $\omega_0$. As $\theta_g \rightarrow 0$, this peaks shifts towards $\omega_0$, which is expected for the case of TLS coupling to a single bath. Thus the density of states can by itself reveal the frustrating effects of decoherence more elegantly.

For the particular case of $H_S=0$ ($\omega_0=0$), one can obtain simplified expressions for the density of states. We obtain
\begin{eqnarray}
D(\omega)&\simeq& \frac{1}{g\sqrt{N}}{\rm e}^{-\omega^2/g^2N} \ ,\quad g_1 = g, \, g_2 = 0 \nonumber \\
D(\omega)&\simeq& \frac{\omega}{g^2N}{\rm e}^{-\omega^2/g^2N} \ , \quad g_1 = g_2 = g
\end{eqnarray}
In the case of a single bath the peak is at $\omega = 0$, where as for two baths, the peak is shifted to $\omega= \sqrt{N/2}g$. We note in this respect that, if the two baths were coupled to same spin component, then the behavior would be similar to that of an effective single-bath coupled to one spin component. In such a case the peak in $D(\omega)$ would remain at $\omega = 0$. Thus, the emergency of a peak at $\omega\neq 0$ may be viewed as a frustration of dissipation due to the competition between two environments coupled to non-commuting spin components.

We end by noting that $D(\omega)$ is the Fourier transform of $\langle S^z(t) S^z(0)\rangle$ and is a measure of the strength of the transitions induced by a periodic perturbation $\sim \cos (\omega t)$.


\section{Correlation functions}

We now investigate the spin--spin correlation functions defined by
${\mathcal C}_{ij}(t) = \langle S^i(t)S^j(0)\rangle$.
Specifically, we focus on its symmetrized and antisymmetrized versions,
 $ {\mathcal S}_{ij}(t) \equiv \frac{1}{2}\langle \{S^i(t),S^j(0)\}\rangle $ and
 $ {\mathcal A}_{ij}(t) \equiv -i\langle [S^i(t),S^j(0)]\rangle$.

Since the system is initially in an eigenstate of $S^z$, the symmetrized autocorrelation function in $z$--direction is simply ${\mathcal S}_{zz}(t) =  -\frac{1}{2} \langle S_z(t)\rangle$. Using the general formulae of Sec.~\ref{secgen}, we find for the transversal symmetrized auto--correlation functions
\begin{equation}
\label{symmcorr}
 {\mathcal S}_{xx}(t) =  \frac{-1}{2^{2N+2}}\sum_{m_1,m_2=-N/2}^{N/2}\lambda_{m_1}\lambda_{m_2}
   \left(\frac{\omega_0^2+g_2^2m_2^2}{\Omega^2_{m_1m_2}}\cos(\Omega_{m_1m_2}t)+ \frac{g_1^2m_1^2}{\Omega_{m_1m_2}^2}\right)
   \end{equation}
and a similar result is obtained for ${\mathcal S}_{yy}(t)$ by exchanging indices $m_1$ and $m_2$.
All symmetrized cross--correlation functions  ${\mathcal S}_{ij}(t)$, ($i\neq j$) are zero.

We look at the autocorrelation function in $z$--direction in the case that there is no external magnetic field applied $\omega_0=0$. Since ${\mathcal S}_{zz}(t)$ is proportional $\langle S^z(t) \rangle$ we can use the integral representation (\ref{ct}).
In general, i.~e. for intermediate values of $\theta_g$ and for non--zero frequency $\omega_0$, the integral~(\ref{ct}) becomes quite difficult and cannot be solved analytically.  However in some limits closed expressions can be derived. For $\omega_0=0$ we obtain
\begin{eqnarray}
\mathcal{S}_{zz}(t) &=& \exp(-Ng^2t^2/8) \ , \quad \theta_g = 0 \nonumber \\
\mathcal{S}_{zz}(t) &=& 1-\sqrt{\frac{\pi Ng^2}{8}}t\exp(-Ng^2t^2/8){\rm Erfi}\left(\frac{\sqrt{N}gt}{2\sqrt{2}}\right)\ , \quad \theta_g = \pi/4
\end{eqnarray}

The antisymmetrized correlation functions are related to the dynamical susceptibilities, defined as
\begin{eqnarray}
\chi_{ij}(\omega) &=&\int_0^\infty \frac{dt}{2\pi}{\rm e}^{i\omega t}{\mathcal A}_{ij}(t) \ .
\end{eqnarray}
In particular, the imaginary part of the susceptibility can be used as a measure of the energy dissipated from the system to the bath. Using Eqs.~(\ref{corr1}) and (\ref{expgen1})
we find ${\mathcal A}_{zz}(t) = 0$,
\begin{eqnarray}
\label{trans1}
{\mathcal A}_{xx}(t) & = &
\frac{-1}{2^{2N+1}}\sum_{m_1,m_2=-N/2}^{N/2}\lambda_{m_1}\lambda_{m_2}
\frac{\omega_0}{\Omega_{m_1m_2}}\sin(\Omega_{m_1m_2}t) \nonumber \\
& = &  \int d\omega D(\omega) \frac{\omega_0}{\omega}\sin(\omega t)
\end{eqnarray}
and ${\mathcal A}_{yy}(t)={\mathcal A}_{xx}(t)$, where $D(\omega)$ has been analyzed in the previous section. All anti--symmetrized cross--correlation function but ${\mathcal A}_{xy}(t)$ are zero. For ${\mathcal A}_{xy}(t)$ we find
\begin{eqnarray}
\label{trans2}
{\mathcal A}_{xy}(t) &=&  \frac{-1}{2^{2N+1}}\sum_{m_1,m_2=-N/2}^{N/2}\lambda_{m_1}\lambda_{m_2}\cos(\Omega_{m_1m_2}t)\nonumber\\
&=& \int d\omega D(\omega)\cos(\omega t)
\end{eqnarray}
In the second lines of Eqs.~(\ref{trans1}) and (\ref{trans2}) we used an integral representation in terms of $D(\omega)$. In this form the dynamical susceptibilities are readily evaluated
\begin{eqnarray}
\chi_{xx}(\omega) = \frac{\omega_0}{2\pi}\int d\omega^\prime \frac{D(\omega^\prime)}{\omega^2-\omega^{\prime 2}+
                         {\rm sgn}(\omega)i0^+}
\end{eqnarray}
Splitting  $\chi_{ij}(\omega)$ in its real and its imaginary part, $\chi_{ij}(\omega) = \chi_{ij}^\prime(\omega) + i\chi_{ij}^{\prime\prime}(\omega)$, one obtains the relation
(\ref{densstate1})
\begin{equation}
\label{susdensrel}
\chi_{xx}^{\prime\prime}(\omega) \ = \ \frac{\omega_0}{2\omega} D(\omega) ,
\end{equation}
which holds for $\omega\geq \omega_0$. Moreover $\chi_{xx}^{\prime\prime}(\omega) = 0$ for $|\omega|< \omega_0$, and $\chi_{xx}^{\prime\prime}(-\omega) = -\chi_{xx}^{\prime\prime}(\omega)$.

We can use the approximation (\ref{c}) and obtain
\begin{eqnarray}
 \chi_{xx}^{\prime\prime}(\omega) & = & \frac{2 \omega_0}{Ng^2}
                          \exp \left(-\frac{4(\omega^2-\omega_0^2)}{Ng^2}\right) \ , \quad \theta_g=\frac{\pi}{4}\\
 \chi_{xx}^{\prime\prime}(\omega) & = & \frac{\omega_0}{\sqrt{2\pi Ng^2(\omega^2-\omega_0^2)}}
                          \exp \left(-\frac{2(\omega^2-\omega_0^2)}{Ng^2}\right) \ , \quad \theta_g= 0
\end{eqnarray}
From the above equations it can be seen that there is a strong singularity at $\omega = \omega_0$, in addition to the Gaussian spread arising due to the interaction with the bath. For the symmetric coupling, this singularity is removed and only a Gaussian spread peaked at $\omega = \omega_0$ remains. Both functions are peaked at $\omega = \omega_0$, and hence one can say that, since there is no peak shifting there is no frustration. If we try to remove the singularity for the single bath case by multiplying $\sqrt{\omega^2-\omega_0^2}$ with $\chi_{xx}^{\prime\prime}(\omega) $ then one can immediately see that the peak for $\chi_{xx}^{\prime\prime}(\omega) $ is shifted away from $\omega = \omega_0$ for the symmetric case. Scalings of such kind can be avoided by considering other kinds of distributions for the bath spins. In the next section we consider bath spin distributions $\lambda_m$ with a Gaussian cutoff.

\section{Tailoring the density of states}

In Refs.~\cite{cas03,nov05} respectively \cite{koh04,koh06b} similar expressions were obtained for $\chi_{xx}^{\prime\prime}(\omega)$ in the first case and for the Fourier transform of the antisymmetrized position--position correlation function in the second case. In both cases the function under consideration is essentially a Lorentzian
\begin{equation}
\label{ds1}
\chi_{xx}^{\prime\prime}(\omega) = \frac{Z\omega}{\left(\omega^2-\widetilde{\omega}_0^2\right)^2+g_{\rm eff}\omega^2} \ ,
\end{equation}
where $\widetilde{\omega}_0$ is the renormalized frequency of the system (Larmor frequency, respectively oscillator frequency), $g_{\rm eff}$ is the effective damping coefficient. For the detailed expressions of $\widetilde{\omega}_0$, $g_{\rm eff}$ and $Z$ see Eq.~(26) of Ref.~\cite{nov05} and Eq.~(20) of Ref.~\cite{koh06b}. The Lorentzian form, and in particular the linear behavior  for small values of $\omega$ is typical for Ohmic type of dissipation.

We model our system to best mimic Ohmic behavior. Since $\chi_{xx}^{\prime\prime}(\omega)$ is related to the density of states $D(\omega)$ by the simple relation (\ref{susdensrel}), we can focus directly on $D(\omega)$ as defined in Eq.~(\ref{densstate}).
We now use a general $\lambda_{m}$ of the form
\begin{equation}
\label{lamgen}
\lambda^{(\alpha)}_m \ =\ \frac{1}{2}\left(\frac{2}{N}\right)^{(\alpha+1)/2}\frac{|m|^\alpha}{\Gamma((1+\alpha)/2)}\exp\left(-\frac{2m^2}{N}\right) \ .
\end{equation}
For small values of $m$ the behavior of $\lambda_m$ is dominated by the power $m^\alpha$ with a characteristic exponent $\alpha$. We note that the case $\alpha =1$ cannot directly be identified with an Ohmic bath. In Eq.~(\ref{lamgen}) we took a Gaussian cutoff for large values of $\omega$ with a cutoff frequency chosen as $\sqrt{N/2}$ in order to make contact with the former results. It is a well known fact in the theory of open quantum systems that the specific form of the cutoff function is not relevant \cite{wei99}. The function $\lambda_m$ is normalized such that $\int dm \lambda(m) = 1$ holds. We see that the form of $\lambda_m$ described in Eq.~(\ref{lambda}) is just a special case of Eq.~(\ref{lamgen}) corresponding to $\alpha=0$ [see also Eq. (\ref{c})].

\begin{figure}[tbp]
\begin{center}
\includegraphics[width=12.0cm]{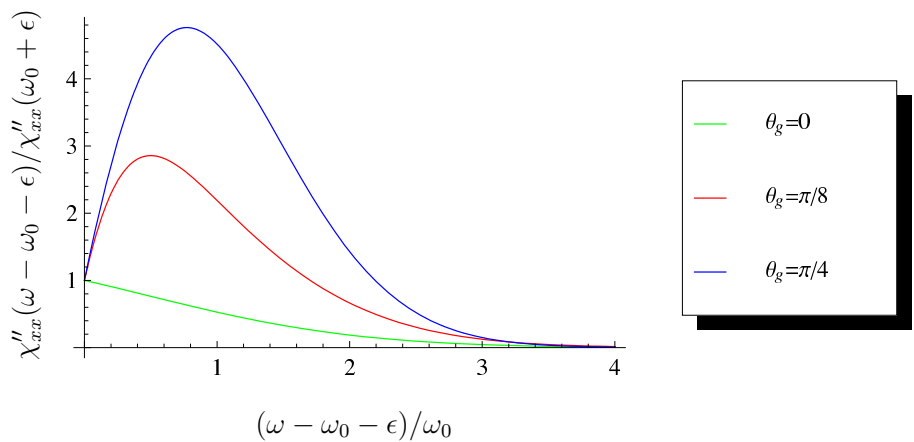}
\caption{Plot of the transverse suceptibility $\chi_{xx}^{\prime\prime}(\omega)$ for an Ohmic type of density of states $\alpha=1$ and for three different angles $\theta_g= 0, \pi/8,\pi/4$.
The other parameters are $\omega_0= 10$, $g=1$ and $N=1000$. The offset at which the function is normalized is $\epsilon=0.1$\label{chiplot1}}\end{center}
\end{figure}

In a calculation which is similar to that performed in Sec. 4, we obtain for the density of states
\begin{eqnarray}
\label{Agen}
D(\omega) &=& \frac{4|\omega|\sqrt{\pi}}{\Gamma((\alpha+1)/2)Ng^2\sin(2\theta_g)}
\left(\frac{2}{Ng^2\cos(2\theta_g)}\right)^{\alpha/2}
               (\omega^2-\omega_0^2)^{\alpha/2}\theta(\omega-\omega_0)\nonumber\\
       && I_{\alpha/2}\left(\frac{4\cot2\theta_g(\omega^2-\omega_0^2)}{Ng^2\sin(2\theta_g)}\right)
 \exp \left(-\frac{4(\omega^2-\omega_0^2)}{Ng^2\sin^2(2\theta_g)}\right) \ ,
\end{eqnarray}
where, as before, $g^2= g_1^2+g_2^2$ and the angle $\theta_g$ is defined in Eq.~(\ref{theta}). Moreover we have introduced the modified
Bessel function $I_{\alpha/2}$ of order $\alpha/2$. For the transverse susceptibility we find in the two limiting cases $\theta_g=0$ and $\theta_g=\pi/4$ the expressions
\begin{eqnarray}
\chi_{xx}^{\prime\prime}(\omega)&=& \frac{1}{2}\frac{\left(\omega^2-\omega_0^2\right)^\alpha\omega_0}
                                  {\Gamma(\alpha+1)}\left(\frac{4}{Ng^2}\right)^{\alpha+1}\nonumber\\
                                &&\quad \exp\left(-\frac{4(\omega^2-\omega_0^2)}{Ng^2}\right)\ ,\quad \theta_g = \frac{\pi}{4}\\
\chi_{xx}^{\prime\prime}(\omega)&=& \frac{1}{2Ng^2}\frac{\left(\omega^2-\omega_0^2\right)^{(\alpha-1)/2}\omega_0}
{\Gamma((\alpha+1)/2)}\left(\frac{2}{Ng^2}\right)^{(\alpha-1)/2}\nonumber\\
                     &&\quad \exp\left(-\frac{2(\omega^2-\omega_0^2)}{N g^2}\right)\ ,\quad \theta_g = 0 \ .
\end{eqnarray}
which satisfies the general properties given after Eq. (\ref{susdensrel}).
The transverse susceptibility is zero for $\omega = \omega_0$. This zero value is found because the distributions $\lambda_m$ is centred at $m=0$. If the distribution is shifted to be centred at a non-zero $m$ value, then the value of $\chi^{\prime\prime}(\omega)_{12}$ will be non-zero at $\omega = \omega_0$ for sufficiently large $\alpha$.
In Fig.~\ref{chiplot1} we have plotted the transverse susceptibility for the symmetric ($\theta_g=\pi/4$) and single-bath ($\theta_g= 0$) cases, as well as for an intermediate situation. In order to compare with other results, in particular with the curves obtained in Ref. \cite{nov05}, we have normalized $\chi_{xx}^{\prime\prime}(\omega)$ so that $\chi_{xx}^{\prime\prime}(\omega_0+\epsilon) = 1$, where $\epsilon$ is a convenient small offset which is chosen for proper scaling and comparison. We note that making $\epsilon$ too small shoots the peak to infinity in those cases where $\lim_{\epsilon \rightarrow 0}\chi_{xx}^{\prime\prime}(\omega_0+\epsilon) = 0$.
Though the natural sum rule $\int d\omega \, \omega \chi_{xx}^{\prime\prime}(\omega) =\omega_0$ is spoiled by adding this epsilon the essential physics behind is unaffected. When there is no interaction with the bath, $\chi^{\prime\prime}(\omega)$ is a delta function peaked at $\omega = \omega_0$.
In the presence of one bath the peak broadens with the maximum still located at $\omega = \omega_0$.

Surprisingly, the peak at $\omega = \omega_0$ disappears when we introduce a second bath which couples to a different component of the central spin. A similar shift was reported in Refs. \cite{cas03,nov05}. It results from a pure frustration effect due to the non-commuting nature of the spin operators.

\begin{figure}[tbp]
\begin{center}
\includegraphics[width=12.0cm]{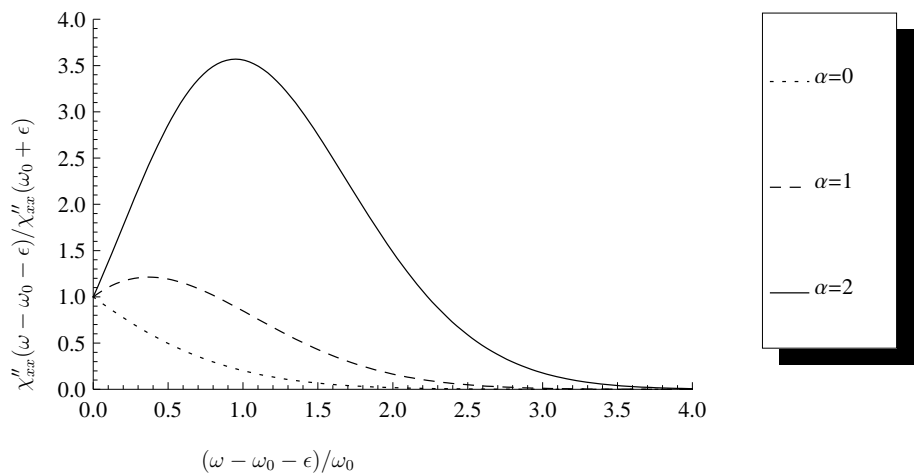}
\caption{Plot of the transverse suceptibility $\chi_{xx}^{\prime\prime}(\omega)$ for the symmetric case $\theta_g= \pi/4$ for the values $\alpha=0$ (dotted), $\alpha=1$ (dashed) and $\alpha=2$ (full). The parameters are $\omega_0= 10$, $g=1$ and $N=1000$. The offset at which the function is normalized is $\epsilon=0.5$}\label{chiplot2}
\end{center}
\end{figure}

In Fig.~\ref{chiplot2} we have plotted the  transverse susceptibility for three different types of infrared behavior $\alpha=0,1,2$, mimicking a subohmic, an Ohmic, respectively a superohmic bath. Dissipation decreases as the power $\alpha$ increases, as expected for general dissipative quantum systems \cite{leg87}. On the other hand we see that the frustration effect of an additional bath increases with increasing power $\alpha$ (not shown).

\section{Conclusions}

We have analyzed the spectral properties of a two-level-system which is coupled to one or two dissipative baths through non-commuting observables. A peak in the spectrum at a nonzero frequency reveals the existence of an effective magnetic field experienced by the central spin. We have seen seen that the coupling to a second bath enhances rather than diminishes that effective field and, with it, the dynamics of the central spin. In the extreme case of a zero external field, the nonzero field is generated solely from the competition of two environments coupling to non-commuting spin components. This fact is remarkable if one notes that the baths are assumed to be initially unpolarized.

These physical effects arising from the non-commuting nature of the spin operators are a general feature of the dynamics which does not depend on details such as the Markovian or non-Markovian character of the reduced system dynamics or the strength of the system-bath interactions. In fact, we explicitly prove that the emergence of a peak in the spectral function is compatible with a fast decay of the quantum impurity. This suggests that, while dissipation is inhibited by the competition between the two baths, decoherence is not, at least in the long time behavior. Our present results are entirely consistent with the results of Refs. \cite{koh05,koh06b} for a harmonic oscillator (equivalent to a large or quasiclassical spin), where a fast decay of the quantum purity was found to coexist with a weak form of suppression of dissipation. Here we have proved that the phenomenon of frustration of dissipation also exists for a dissipative two-level-system which, given its reduced dimensionality, is much more quantum in nature than the harmonic oscillator.

The intuitive idea that two competing environments attempt to measure non-commuting observables and thus fail to generate decoherence sounds appealing but may be misleading. The statement would be true if the only possible result of a quantum measurement were to select a narrow distribution of eigenstates of the measured observable, since two non-commuting observables cannot be simultaneously well defined. However, a possible outcome of the coupling to a dissipative environment is that the reduced density matrix, while becoming diagonal in the representation of the eigenstates of the measured observable, may display a broad probability distribution in that representation. In the limit in which that distribution is very broad, the reduced density approaches the identity matrix, which is invariant under a change of basis. Thus a reduced density matrix may be simultaneously diagonal in the representations of two non-commuting observables, provided it is close to the identity matrix. This is what actually happens to our central spin-$\frac{1}{2}$, as is clearly revealed by the quantum purity tending to its minimum value 1/2 at long times, both for a single and a symmetric double bath. As we have seen, this feature is compatible with the reinforcement of the central spin dynamics resulting from the competition of the two environments. The upshot of the present study on the effect of competing environments is that, at least for dissipative two-level-systems, it may be misleading to speak of quantum frustration of decoherence and is more appropriate to introduce the concept of quantum frustration of dissipation.

We cannot rule out however the possibility that the coexistence of decoherence and dynamics enhancement by two competing environments is a particular property of our dissipative model where the bath has no internal dynamics. A firmer conclusion on the existence or absence of decoherence frustration will require an understanding of the behavior of genuinely quantum properties such as purity or pair entanglement in the presence of competing environments with internal dynamics.

\ack This work has been supported by the German Research Council (DFG)
through grant No. Ko 3538/1-1 and by MEC (Spain) through grants No.
FIS2004-05120 and FIS2007-65723.

\section*{References}
\bibliographystyle{plain}

\begin{thebibliography}{10}

\bibitem{leg87}
A.~J. Leggett, S.~Chakravarty, A.~T. Dorsey, M.~P.~A. Fisher, A.~Garg, and
  W.~Zwerger.
\newblock {\em Rev. Mod. Phys.}, 59:1, 1987.

\bibitem{wei99}
U.~Weiss.
\newblock {\em Quantum Dissipative Systems}.
\newblock World Scientific, Singapore, 2nd edition, 1999.

\bibitem{cas03}
A.~H.~Castro Neto, E.~Novais, L.~Borda, G.~Zarand, and I.~Affleck.
\newblock {\em Phys. Rev. Lett.}, 91:096401, 2003.

\bibitem{nov05}
E.~Novais, A.~H.~Castro Neto, L.~Borda, I.~Affleck, and G.~Zarand.
\newblock {\em Phys. Rev. B}, 72:014417, 2005.

\bibitem{koh04}
H.~Kohler, F. Guinea and F.~Sols.
\newblock {\em Ann. Phys.}, 72:014417, 2004.

\bibitem{giu07}
D. Giuliano and P. Sodano.
\newblock arXiv:0710.5554.

\bibitem{koh05}
H.~Kohler and F.~Sols.
\newblock {\em Phys. Rev. B}, 72:014417, 2005.

\bibitem{koh06b}
H.~Kohler and F.~Sols.
\newblock {\em New J. Phys.}, 8:149, 2006.

\bibitem{pro00}
N.~V. Prokof'ev and P.~C.~E. Stamp.
\newblock {\em Rep. Prog. Phys.}, 63:669, 2000.

\bibitem{cuc05}
F.~M. Cucchietti, J.~P. Paz, and W.~H. Zurek.
\newblock {\em Phys. Rev. B}, 70:035311, 2005.


\bibitem{rel07}
A. Relano, J. Dukelsky, and R. A. Molina.
\newblock {\em Phys. Rev. E}, 76:046223, 2007.


\bibitem{zur07}
W.~H. Zurek, F.~M. Cucchietti, and J.~P. Paz.
\newblock {\em Acta Phys. Polonica}, 38:1685, 2005.


\bibitem{ros07a}
D. Rossini, T. Calarco, V. Giovannetti, S. Montangero, and  R. Fazio.
\newblock {\em Phys. Rev. A}, 75:032333, 2007.

\bibitem{ros07b}
D. Rossini, T. Calarco, V. Giovannetti, S. Montangero, and  R. Fazio.
\newblock {\em J. Phys A: Math. Theor.}, 40:8033, 2007.

\bibitem{rao06}
D.~D.~Bhaktavatsala Rao, V. Ravishankar and V. Subrahmanyam.
\newblock {\em Phys. Rev. A}, 74:22301, 2006.

\bibitem{rao07}
D.~D.~Bhaktavatsala Rao.
\newblock {\em Phys. Rev. A}, 76:042312, 2007.

\bibitem{lai08}
C. Y.  Lai, J. T. Hung, C. Y. Mou, P. Chen.
\newblock {\em Phys. Rev. B}, 77:205419, 2008.

\bibitem{loss}
J. Schliemann, A. Khaetskii, and D. Loss.
\newblock {\em J. Phys.: Cond. Matter}, 15:1809, 2003.

\bibitem{gri01}
G.~R. Grimmett and D.~R. Stirzaker.
\newblock {\em Probability and Random Processes}.
\newblock Oxford University Press, Oxford, 3nd edition, 2001.


\end{thebibliography}

\end{document}